\documentclass[aps,amsmath,amssymb,nofootinbib]{revtex4}
\everymath{\displaystyle}
\usepackage{graphicx}

\begin{document}
\title{Thermodynamics of strong coupling 2-color QCD \\
       with chiral and diquark condensates}
\author{Y.~Nishida, K.~Fukushima, and T.~Hatsuda}
\affiliation{Department of Physics, University of Tokyo,
             7-3-1 Hongo, Bunkyo-ku, Tokyo 113-0033, Japan}
\date{\today}
\begin{abstract}
 2-color QCD (quantum chromodynamics with $N_{\mathrm{c}}=2$) at
 finite temperature $T$ and chemical potential $\mu$ is revisited in
 the strong coupling limit on the lattice with staggered
 fermions. The phase structure in the space of $T$, $\mu$, and the
 quark mass $m$ is elucidated with the use of the mean field
  approximation and the $1/d$ expansion ($d$ being the number of
 spatial dimensions). We put special emphasis on the interplay among
 the chiral condensate $\langle\bar{q}q\rangle$, the diquark
 condensate $\langle qq\rangle$, and the quark density
 $\langle q^{\dagger}q\rangle$ in the $T$-$\mu$-$m$ space. Simple
 analytic formulae are also derived without assuming $\mu$ nor $m$
 being small. Qualitative comparisons are made between our results
 and those of recent Monte-Carlo simulations in 2-color QCD.
\end{abstract}
\maketitle


\section{INTRODUCTION}

Physics of matter under high baryon density is one of the most
challenging problems in quantum chromodynamics (QCD) both from
technical and physical point of view. So far there have been proposed
various novel phases which include the $^3\mathrm{P}_2$ neutron
superfluidity \cite{TAMA70}, the pion \cite{MIG78} and kaon
\cite{KN86} condensations, the deconfined quark matter \cite{CP75},
the strange matter \cite{witten}, the color superconductivity
\cite{BL84}, the ferromagnetic quark ordering \cite{tatsumi} and so
on.

Unfortunately, analyzing these phases from the first principle lattice
simulations is retarded due to the complex fermion determinant for
3-color QCD at finite baryon chemical potential with quarks in the
fundamental representation. The situation is however different for
2-color QCD in which the fermion determinant can be made real and the
Monte-Carlo simulations are attainable as was first demonstrated in
\cite{NAKA84}. Because of this reason, 2-color QCD provides an unique
opportunity to compare various ideas at finite chemical potential with
 the results from lattice simulations.

One of the major differences between 2-color QCD and 3-color QCD lies
in the fact that the color-singlet baryons are bosons in the former.
This implies that the ground state of the 2-color system at finite
baryon density in the color-confined phase is an interacting boson
system, i.e.\ a Bose liquid, although the quark Fermi liquid may be
realized at high baryon density in the color-deconfined phase. How
this Bose liquid changes its character as a function of the
temperature $T$, the quark chemical potential $\mu$, and the quark
mass $m$ is an interesting question by itself and may also give a
hint to understand physics of the color superconductivity in 3-color
QCD in which the crossover from the Bose-Einstein condensate of
tightly bound quark pairs to the BCS type condensate of loosely bound
Cooper pairs may take place \cite{itakura02}.

In the present paper, we revisit the thermodynamics of the strong
coupling limit of 2-color lattice QCD with staggered fermions with
chiral and diquark condensates which was originally studied in
\cite{DKM84,DMW87,KM90}. Our main purpose is to analyze its phase
structure and the interplay among the chiral condensate
$\langle\bar{q}q\rangle$, the diquark condensate $\langle qq\rangle$,
and the baryon density $\langle q^{\dagger}q\rangle$ as functions of
$T$, $\mu$, and $m$. This would give not only a useful guide to the
actively pursued lattice QCD simulations of the same system in the
weak coupling \cite{han99,kog02,nonaka03} but also give us physical
insight into the Bose liquid together with the other knowledge from
the instanton liquid model \cite{instanton}, the random matrix model
\cite{RMT}, the chiral perturbation theory \cite{ChPT0,ChPT}, and the
renormalization group \cite{RGE}.

This paper is organized as follows. In Sec.~\ref{sec:formulation}, we
introduce 2-color QCD with one-component staggered quarks (which
correspond to quarks with 4 flavors in the continuum limit). By
introducing the auxiliary fields $\sigma$ and $\Delta$ corresponding
to the chiral and diquark condensates, we derive an effective free
energy $F_{\mathrm{eff}}[\sigma,\Delta]$ in the $1/d$ expansion ($d$
being the number of spatial dimensions) and the mean field
approximation. The resultant free energy is enough simple so that one
can make analytic studies at least in the chiral limit $m=0$ with
finite $T$ and $\mu$, and in the zero temperature limit $T=0$ with
finite $\mu$ and $m$. Sec.~\ref{sec:analytic} is devoted to such
studies and useful relations of the critical temperature and chemical
potential are derived. In Sec.~\ref{sec:numerical}, we make numerical
analyses on the chiral and diquark condensates as well as the quark
number density as functions of $T$, $\mu$, and $m$. The phase
structure of 2-color QCD at strong coupling is clarified in the three
dimensional $(T, \mu, m)$ space. In Sec.~\ref{sec:summary}, summary
and concluding remarks are given. In Appendices \ref{app:summation}
and \ref{app:proof}, we give some technical details in deriving and
analyzing the free energy.


\section{FORMULATION}
\label{sec:formulation}

In this section, we derive an effective action for meson and diquark
fields starting from the lattice action with the staggered fermion. The
readers who are interested in the results in advance can skip the
derivation and directly refer to the expression of the free energy
given in Eq.~(\ref{eq:free_energy}) with Eq.~(\ref{eq:quasi_energy}).

The action on the lattice is given by
\begin{align}
 S[U,\chi,\bar{\chi}]=S_{\mathrm{G}}[U]
  +S_{\mathrm{F}}[U,\chi,\bar{\chi}]\,,
\label{eq:action}
\end{align}
which consists of the gluonic part,
\begin{align}
 S_{\mathrm{G}}[U]=\frac{2N_{\mathrm{c}}}{g^2}\sum_{x,\mu,\nu}
  \left\{1-\frac{1}{N_{\mathrm{c}}}\mathrm{Re}\mathrm{Tr}\,
  U_{\mathrm{\mu\nu}}(x)\right\}\,, \qquad
 U_{\mu\nu}(x)=U_\nu^\dagger(x)U_\mu^\dagger(x+\hat{\nu})
  U_\nu(x+\hat{\mu})U_\mu(x)\,,
\label{eq:action_gluon}
\end{align}
and the fermionic part with finite chemical potential \cite{HK83},
\begin{align}
 \begin{split}
  S_{\mathrm{F}}[U,\chi,\bar{\chi}]=m\sum_x\bar{\chi}(x)\chi(x)
   &+\frac{1}{2}\sum_x\sum_{j=1}^d\eta_j(x)
   \left\{\bar{\chi}(x)U_j(x)\chi(x+\hat{j})
   -\bar{\chi}(x+\hat{j})U_j^\dagger(x)\chi(x)\right\}\\
   &+\frac{1}{2}\sum_x\eta_0(x)\left\{\bar{\chi}(x)\mathrm{e}^\mu
   U_0(x)\chi(x+\hat{0})-\bar{\chi}(x+\hat{0})U_0^\dagger(x)
   \mathrm{e}^{-\mu}\chi(x)\right\}\,.
 \end{split}
  \label{eq:action_quark}
\end{align}
$\chi$ stands for the quark field in the fundamental representation
of the color $\mathrm{SU}(N_{\mathrm{c}})$ group and $U_\mu$ is the
$\mathrm{SU}(N_{\mathrm{c}})$ valued gauge link variable. $g$ is the
gauge coupling constant and $d$ represents the number of spatial
directions which takes 3 in reality. Later we sometimes use a notation
$x=(\tau,\vec{x})$ in which $\tau$ ($\vec{x}$) represents the temporal
(spatial) coordinate. $\eta_0(x)$ and $\eta_j(x)$ inherent in the
staggered formalism are defined as
\begin{align}
 \eta_0(x)=1\,,\qquad\eta_j(x)=(-1)^{\sum_{i=1}^j x_{i-1}}\,.
\end{align}
$\mu$ in Eq.~(\ref{eq:action_quark}) is the quark chemical potential,
while the temperature $T=(aN_\tau)^{-1}$ with $a$ being the lattice
spacing and $N_\tau$ being the number of temporal sites. We will write
all the dimensionful quantities in unit of $a$ and will not write $a$
explicitly. Since we are interested in 2-color QCD in this paper, we
take $N_{\mathrm{c}}=2$ in the following.

It is worth mentioning here on the symmetry breaking pattern of
2-color QCD formulated on the lattice. The staggered fermion with a
single component, which corresponds to $N_{\mathrm{f}}=4$ in the
continuum limit, has the global $\mathrm{U_V(1)\times U_A(1)}$
symmetry for $m=\mu=0$ \cite{klu83}. Here $\mathrm{U_V(1)}$
($\mathrm{U_A(1)}$) corresponds to the conservation of the baryon
number (axial charge). For $N_{\mathrm{c}}=2$,
$\mathrm{U_V(1)\times U_A(1)}$ is graded to a larger symmetry U(2)
\cite{han99,klu83}. This is because the color SU(2) group is
pseudo-real and the action possesses Pauli-G\"{u}rsey's
fermion$-$anti-fermion symmetry \cite{PG57}.

Introduction of a finite chemical potential $\mu$ explicitly breaks
the U(2) symmetry down to the original $\mathrm{U_V(1)\times U_A(1)}$.
Further introduction of a finite quark mass $m$ retains only the
$\mathrm{U_V(1)}$ symmetry. Listed in Table~\ref{tab:symmetry} are the
symmetries realized in various circumstances and their breaking
patterns.

\begin{table}[htbp]
\begin{tabular}{|c|c|c|}
\hline
 & $m=0$ & $m\neq0$ \\ \hline
$\mu=0$ & U(2) & $\mathrm{U_V(1)}$ \\
 & broken to U(1) with 3 NG modes & not broken \\ \hline
$\mu\neq0$ & $\mathrm{U_V(1)\times U_A(1)}$ & $\mathrm{U_V(1)}$ \\
 & totally broken with 2 NG modes & totally broken with 1 NG mode \\
\hline
\end{tabular}
\caption{Symmetry realized  in the single component staggered-fermion
action for $N_{\mathrm{c}}=2$. Possible symmetry breaking patterns and
the number of Nambu-Goldstone (NG) modes are also listed.}
\label{tab:symmetry}
\end{table}
 
Before addressing the computational details to derive the effective
action for the meson and diquark system from the original action
Eq.~(\ref{eq:action}), we shall summarize the actual procedure which
is similar to that of \cite{DMW87}:

\begin{description}

\item{[Step 1]} Strong coupling limit $g\rightarrow\infty$ is taken.
 Then the gluonic part of the action $S_{\mathrm{G}}[U]$ in
 Eq.~(\ref{eq:action_gluon}) vanishes, because it is inversely
 proportional to $g^2$. Consequently the gauge field remains only in
 the fermionic part (\ref{eq:action_quark}).

\item{[Step 2]} Large dimensional ($1/d$) expansion is employed in the
 spatial directions in order to facilitate the integration
 over the spatial link variable $U_j$. The temporal link variable
 $U_0$ is left untouched at this Step and will be exactly integrated
 out later in Step 4.

\item{[Step 3]} Bosonization is performed by introducing the auxiliary
 fields $\sigma$ for $\bar{\chi}\chi$ and $\Delta$ for $\chi\chi$.
 Then the mean field approximation is adopted for the auxiliary
 fields. Namely, $\sigma$ and $\Delta$ are regarded as spatially
 uniform condensates.

\item{[Step 4]} Integration with respect to $\chi$, $\bar{\chi}$, and
 $U_0$ are accomplished exactly to result in an effective free energy
 written in terms of $\sigma$ (chiral condensate) and $\Delta$
 (diquark condensate).

\end{description}

Some comments are in order here to emphasize the importance of the
exact integration over $U_0$ without using the $1/d$ expansion in Step
4. Suppose we performed the $1/d$ expansion not only in the spatial
directions  but also in the temporal direction to perform the $U_j$
and $U_0$ integrations. Then, the quarks would be totally replaced by
the mesons and diquarks, and the temperature would not enter into the
free energy within the mean field approximation. In other words, one
needs to consider meson and diquark fluctuations to obtain thermal
effects in such an approach. In our procedure, on the other hand, the
thermal effect can be incorporated even in the mean field level by
exempting the $1/d$ expansion in making $U_0$ integration. Similar
observation has been made also in an exactly solvable many fermion
system \cite{hatsuda89}, in which it was shown that the approach
similar to ours has better convergence to the exact result.

Let us now describe each procedure in some details with a particular
emphasis on the importance of the interplay between the chiral and
diquark condensates.


\subsection{Strong coupling limit and the $1/d$ expansion (Step 1 and 2)}

After taking the strong coupling limit $g\to\infty$, the partition
function is written as
\begin{align}
 Z=\int\mathcal{D}[\chi,\bar{\chi}]\,\mathcal{D}[U_0]\,
  \mathcal{D}[U_j]\,\mathrm{e}^{-S_{\mathrm{F}}[U,\chi,\bar{\chi}]}\,.
\end{align}
Because $\chi$ and $\bar{\chi}$ are fermion fields, the Taylor
expansion of $\mathrm{e}^{-S_{\mathrm{F}}[U,\chi,\bar{\chi}]}$
generates at most $2^{2N_{\mathrm{c}}}$ terms on each site $x$. Thus
the integration with respect to the link variable could be performed
exactly in principle. Instead, we truncate such an expansion here up
to the leading order in the $1/d$ expansion. As explained soon below,
the lowest order of the Taylor expansion gives the leading
contribution in the $1/d$ expansion. Then we can integrate each term
with respect to $U_j$ at each site $x$. As we stated in Step 2 of the
above summary, $U_0$ is left untouched at this stage:
\begin{align}
 &\int\mathrm{d}[U_j(x)]\,\exp\left[-\frac{1}{2}
  \sum_{j=1}^d \eta_j(x)\left\{\bar{\chi}(x)U_j(x)\chi(x+\hat{j})
  -\bar{\chi}(x+\hat{j})U_j^\dagger(x)\chi(x)\right\}\right]\notag\\
 &=1+\frac{1}{8}\sum_{j=1}^d\bar{\chi}^a(x)\chi^a(x)\bar{\chi}^b
  (x+\hat{j})\chi^b(x+\hat{j})-\frac{1}{16}\sum_{j=1}^d\epsilon_{ab}
  \bar{\chi}^a(x)\bar{\chi}^b(x)\epsilon_{cd}\chi^c(x+\hat{j})
  \chi^d(x+\hat{j})+\text{h.\,c.}+\mathrm{O}(1/d)\notag\\
 &=\exp\left[\sum_y M(x)V_{\mathrm{M}}(x,y)M(y)+\sum_y
  \bar{B}(x)V_{\mathrm{B}}(x,y)B(y)\right]+\mathrm{O}(1/d),
\label{eq:action_MB}
\end{align}
where the Latin indices $a,b,\dots$ are summed over in the color
space. In integrating the link variable, we have utilized the
formulae for the SU(2) group integration,
\begin{align}
 \int\mathrm{d}[U]=1\,,\quad
 \int\mathrm{d}[U]\,U_{ab}U^\dagger_{cd}=\frac{1}{2}
  \delta_{ad}\delta_{bc}\,,\quad
 \int\mathrm{d}[U]\,U_{ab}U_{cd}=\frac{1}{2}\epsilon_{ac}
  \epsilon_{bd}\,.
\end{align}
The mesonic composite $M(x)$ and the baryonic composite $B(x)$ are
defined respectively as
\begin{align}
 M(x)=\frac{1}{2}\delta_{ab}\,\bar{\chi}^a(x)\chi^b(x),\quad
  B(x)=\frac{1}{2}\epsilon_{ab}\,\chi^a(x)\chi^b(x),\quad
  \bar{B}(x)=-\frac{1}{2}\epsilon_{ab}\,\bar\chi^a(x)
  \bar{\chi}^b(x)\,,
\end{align}
and their propagators in the spatial directions are given by
\begin{align}
 V_{\mathrm{M}}(x,y)=V_{\mathrm{B}}(x,y)
  =\frac{1}{4}\sum_{j=1}^d\left(\delta_{y,x+\hat{j}}
  +\delta_{y,x-\hat{j}}\right) .
\end{align}

The reason why the bilinear forms of $M(x)$ and $B(x)$ correspond to
the leading order of the $1/d$ expansion can be understood by changing
the normalization as
$V_{\mathrm{M}}(x,y)/d\to\tilde{V}_{\mathrm{M}}(x,y)$ and
$M(x)\to\tilde{M}(x)/\sqrt{d}$ so that the mesonic propagator
$\tilde{V}_{\mathrm{M}}(x,y)$ is O(1) in the large $d$ limit.
As a result, the more $M(x)$ is contained in the higher order terms in
the Taylor expansion, the more $1/\sqrt{d}$ is associated with it. The
same argument holds exactly for the baryonic composite $B(x)$ with
$N_{\mathrm{c}}=2$, where $B(x)$ is composed of two quarks. For
$N_{\mathrm{c}}>2$, however, the baryonic contribution is of higher
order in the $1/d$ expansion as compared with the mesonic one. For
further details on the $1/d$ expansion for general $N_{\mathrm{c}}$,
see also the discussion in \cite{klu83}.


\subsection{Bosonization and the mean field approximation (Step 3)}

The resultant action (\ref{eq:action_MB}) describes the nearest
neighbor interaction between the mesonic composites and between the
baryonic composites. Since they are the four-fermion interactions,
one may linearize them with a standard Gaussian  technique with
the auxiliary fields $\sigma$ and $\Delta$;
\begin{align}
 &\exp\left[\sum_{x,y}M(x)V_{\mathrm{M}}(x,y)M(y)\right]
  =\int\mathcal{D}[\sigma]\,\exp\left[-\sum_{x,y}\left\{
  \sigma(x)V_{\mathrm{M}}(x,y)\sigma(y)
  +2\sigma(x)V_{\mathrm{M}}(x,y)M(y)\right\}\right]\,,
\intertext{and}
 \begin{split}
  &\exp\left[\sum_{x,y}\bar B(x)V_{\mathrm{B}}(x,y)B(y)\right]\\
  &\qquad\quad=\int\mathcal D[\Delta]\,\exp\left[
   -\sum_{x,y}\left\{\Delta^\ast(x)V_{\mathrm{B}}(x,y)\Delta(y)
   -\Delta^\ast(x)V_\mathrm{B}(x,y)B(y)-\bar B(x)V_{\mathrm{B}}(x,y)
   \Delta(y)\right\}\right]\,.
 \end{split}
\end{align}
 From the above transformations, it is easy to show the relation,
\begin{align}
 \left\langle\sigma(x)\right\rangle=-\left\langle M(x)\right\rangle\,,
  \qquad
 \left\langle\Delta(x)\right\rangle=\left\langle B(x)\right\rangle\,,
  \quad
 \left\langle\Delta^\ast(x)\right\rangle=\left\langle\bar B(x)\right\rangle\,,
\label{eq:condensate}
\end{align}
where we have intentionally chosen the definition of the field
$\sigma$ so that the sign of $\langle\sigma(x)\rangle$ becomes
positive for $m>0$.

Now we take a mean field approximation. Namely, we replace the
auxiliary fields $\sigma(x)$ and $\Delta(x)$ by the constant
condensates $\sigma$ and $\Delta$ and ignore any fluctuations around
the condensates. It is obvious from Eq.~(\ref{eq:condensate}) that
$\sigma$ should be identified with the chiral condensate and $\Delta$
be the diquark condensate. Then we can write the partition function as
\begin{align}
  Z&=\int\mathcal{D}[U_0]\,\mathcal{D}[\chi,\bar{\chi}]\,
  \mathrm{e}^{-S'[U_0,\chi,\bar{\chi};\sigma,\Delta]}\,,
\label{eq:partition}
\intertext{with}
 \begin{split}
  S'[U_0,\chi,\bar{\chi};\sigma,\Delta]
  &=\sum_x\left[m\bar\chi(x)\chi(x)+\frac{1}{2}\left\{\bar\chi(x)
   \mathrm{e}^\mu U_0(x)\chi(x+\hat0)-\bar{\chi}(x+\hat0)
   U_0^\dagger(x)\mathrm{e}^{-\mu}\chi(x)\right\}\right.\\
  &\qquad\qquad\qquad\left.+\frac{d}{2}\sigma^2+\frac{d}{2}|\Delta|^2
   +d\,\sigma M(x)-\frac{d}{2}\Delta^\ast B(x)
   -\frac{d}{2}\Delta\bar{B}(x)\right]\,.
 \end{split}
\label{eq:priact}
\end{align}
Since $S'[U_0,\chi,\bar{\chi};\sigma,\Delta]$ is in a bilinear form
with respect to the quark fields $\chi$ and $\bar{\chi}$, we can
integrate out them immediately. The integration with respect to $U_0$,
which seems to be tough at first glance, turns out to be feasible as
demonstrated in the next subsection.


\subsection{Integrations over $\chi$, $\bar{\chi}$, and $U_0$ (Step 4)}

In order to complete the remaining integrals, we adopt a particular
gauge in which $U_0(\tau,\vec{x})$ is diagonal and independent of
$\tau$ (often called the Polyakov gauge), 
\begin{align}
 U_0(\tau,\vec{x})=\mathrm{diag}
  \left(\mathrm{e}^{\mathrm{i}\theta_1(\vec{x})/N_\tau},\,
  \mathrm{e}^{\mathrm{i}\theta_2(\vec{x})/N_\tau}\right)\,,
   \qquad \text{with} \quad\theta_1(\vec{x})=-\theta_2(\vec{x})\,.
\label{eq:gauge_choice}
\end{align}
Also we make a partial Fourier transformation for the quark fields;
\begin{align}
 \chi(\tau,\vec{x})=\frac{1}{\sqrt{N_\tau}}\sum_{m=1}^{N_\tau}
  \mathrm{e}^{\mathrm{i}k_m\tau}\tilde{\chi}(m,\vec{x})\,,\quad
 \bar{\chi}(\tau,\vec{x})=\frac{1}{\sqrt{N_\tau}}\sum_{m=1}^{N_\tau}
  \mathrm{e}^{-\mathrm{i}k_m\tau}\tilde{\bar{\chi}}(m,\vec{x})\,,
 \qquad k_m=2\pi\frac{(m-\frac{1}{2})}{N_\tau}\,.
\label{eq:fourier}
\end{align}
The anti-periodic condition in the temporal direction,
$\chi(\tau+N_\tau,\vec{x})=-\chi(\tau,\vec{x})$, is satisfied owing to
the fermionic Matsubara frequency $k_m$.

Substituting Eqs.~(\ref{eq:gauge_choice}) and (\ref{eq:fourier}) into
the action (\ref{eq:priact}) and taking the summation over $\tau$, we
reach the action in a Nambu-Gor'kov representation;
\begin{align}
 S'[\theta,\chi,\bar{\chi};\sigma,\Delta]
  =\sum_x\left(\frac{d}{2}\sigma^2+\frac{d}{2}|\Delta|^2\right)
  -\frac{1}{2}\sum_{\vec{x}}\sum_{m,n}\,\!
  [X^a(m,\vec{x})]^{\mathrm{t}}\,G^{-1}_{ab}(m,n;\theta(\vec{x}))\,
  X^b(n,\vec{x})\,,
\end{align}
where
\begin{align}
 X^a(m,\vec x)=
  \begin{pmatrix}
\tilde{\chi}^a(m,\vec{x})\\ \tilde{\bar{\chi}}^a(m,\vec{x})
  \end{pmatrix},
\end{align}
and
\begin{align}
 G^{-1}_{ab}(m,n;\theta(\vec x))=
  \begin{pmatrix}
\frac{d}{2}\Delta^\ast\,\delta_{m,N_\tau-n+1}\epsilon_{ab}
&\left[M+\mathrm{i}\sin\left(k_m+\frac{\theta_a(\vec{x})}
{N_\tau}-\mathrm{i}\mu\right)\right]\delta_{mn}\delta_{ab}\\
-\left[M+\mathrm{i}\sin\left(k_m+\frac{\theta_a(\vec{x})}
{N_\tau}-\mathrm{i}\mu\right)\right]\delta_{mn}\delta_{ab}
&-\frac{d}{2}\Delta\,\delta_{m,N_\tau-n+1}\epsilon_{ab}
  \end{pmatrix}\,.
\end{align}
The indices $a$ and $b$ run from 1 through 2 in the color space. $M$
denotes the dynamical quark mass (to be distinguished from the mesonic
composite field $M(x)$) defined by
\begin{align}
 M=m+\frac{d}{2}\sigma\,.
\label{eq:d_mass}
\end{align}

We can perform the Grassmann integration over $\chi$ and $\bar{\chi}$
as \cite{zin89}
\begin{align}
 \int\mathcal{D}[\chi,\bar{\chi}]\,\mathrm{e}^{X^{\mathrm{t}}G^{-1}X}
  =\prod_{\vec{x}}\sqrt{\mathrm{Det}\left[
  G^{-1}_{ab}(m,n;\theta(\vec{x}))\right]}\,.
\label{eq:grassmann_integral}
\end{align}
Det stands for the determinant with respect to the color indices and
the Matsubara frequencies.  The square root of the determinant  
may be simplified as
\begin{align}
 \sqrt{\mathrm{Det}[G^{-1}]}=\prod_{m=1}^{N_\tau}
  \left[\left(\frac{d}{2}\right)^2|\Delta|^2+\left\{M+\mathrm{i}\sin
  \left(k_m+\theta/N_\tau-\mathrm{i}\mu\right)\right\}\cdot\left\{M
  -\mathrm{i}\sin\left(k_m+\theta/N_\tau+\mathrm{i}\mu\right)\right\}
  \right]\,,
  \label{eq:pre-det}
\end{align}
where $\theta_1=-\theta_2=\theta$ is substituted. The product with
respect to $m$ can be performed using a technique similar to that in
the calculation of the free energy in finite-temperature field theory
in the continuum. The details of the calculation is given in
Appendix~\ref{app:summation}. The result turns out to be a rather
simple form,
  \begin{align}
 \sqrt{\mathrm{Det}\left[G^{-1}_{ab}(m,n;\theta)\right]}
  =\left(\cos\theta+\cosh N_\tau E_-\right)\cdot
  \left(\cos\theta+\cosh N_\tau E_+\right)\,.
\label{eq:sqrtdet}
\end{align}
Here $E_{\pm}$ is the excitation energy of quasi-quarks,
\begin{align}
 E_\pm =\mathrm{arccosh}\left(\sqrt{(1+M^2)\cosh^2\mu
       +(d/2)^2|\Delta|^2}\pm M\sinh\mu\right)
\label{eq:quasi_energy}
\end{align}
with the dynamical quark mass $M$ defined in Eq.~(\ref{eq:d_mass}).

Finally, all we have to do is to integrate this resultant determinant
with respect to $U_0$, or $\theta$, to derive the effective action;
\begin{align}
 S_{\mathrm{eff}}[\sigma,\Delta]=-\log Z
  =\sum_x\left(\frac{d}{2}\sigma^2+\frac{d}{2}|\Delta|^2\right)
  -\sum_{\vec{x}}\log\left\{\int\frac{\mathrm{d}\theta(\vec{x})}{2\pi}
  \sin^2\theta(\vec{x})\sqrt{\mathrm{Det}
  \left[G^{-1}_{ab}(m,n;\theta(\vec{x}))\right]}\right\}\,,
\label{eq:effact}
\end{align}
where we have used the SU(2) Haar measure in the Polyakov gauge
(\ref{eq:gauge_choice}),
\begin{align}
 \int\mathcal{D}[U_0]=\left.\prod_{\vec{x}}\int_{-\pi}^\pi
   \frac{\mathrm{d}\theta(\vec{x})}{2\pi}\sin^2\theta(\vec{x})
   \right|_{\theta(\vec{x})=\theta_1(\vec{x})=-\theta_2(\vec{x})}\,.
\label{eq:measure}
\end{align}
The $U_0$ integration projects out the color singlet pairing of quarks
and anti-quarks among arbitrary excitations in the determinant. Thus
what excites thermally is no longer single quarks, but color singlet
mesons or diquarks. In this sense, the strong coupling limit in
2-color QCD inevitably leads to the boson system.

Eqs.~(\ref{eq:sqrtdet}), (\ref{eq:effact}), and (\ref{eq:measure})
immediately yield the effective free energy;
\begin{align}
 F_{\mathrm{eff}}[\sigma,\Delta]
  =S_{\mathrm{eff}}/({\textstyle \sum_x})=\frac{d}{2}\sigma^2
  +\frac{d}{2}|\Delta|^2-T\log\left\{1+4\cosh\left(E_+/T\right)\cdot
  \cosh\left(E_-/T\right)\right\}\,,
\label{eq:free_energy}
\end{align}
where we have rewritten $N_\tau$ in terms of the temperature
$T(=1/N_\tau)$.

Although the quasi-quark energy $E_\pm$ given in
Eq.~(\ref{eq:quasi_energy}) takes a complicated form, $E_\pm$ is
reduced to a simple expression in the \textit{naive} continuum
$a\to 0$ with $a$ being the lattice spacing. Assuming $Ma$,
$\Delta a$, $\mu a\ll1$ in this limit, $E_\pm$ amounts to a familiar
form in the continuous space-time,
\begin{align}
 E_\pm\sim \mathrm{arccosh}\left(\sqrt{1+M^2+\mu^2
  +(d/2)^2|\Delta|^2}\pm M\mu\right)
  \to\sqrt{(M\pm\mu)^2+(d/2)^2|\Delta|^2}\,.
\end{align}
The quasi-quarks are static in this framework because of the mean
field approximation.


\section{ANALYTIC RESULTS ON THE PHASE STRUCTURE}
\label{sec:analytic}


\subsection{Case in the Chiral Limit}
\label{sec:analytic-A}

In the chiral limit $m=0$ with zero chemical potential $\mu=0$, the
free energy in the mean field approximation given in
Eq.~(\ref{eq:free_energy}) is a function only in terms of
$\sigma^2+|\Delta|^2$. As a result, the free energy is invariant under
the transformation mixing the chiral condensate with the diquark one. 
This corresponds to a subgroup of the U(2) symmetry of the original
action at $m=\mu=0$ in Table~\ref{tab:symmetry}. Because of  this
symmetry, the chiral condensate is indistinguishable from the diquark
condensate for $m=\mu=0$, so that a state with finite $\sigma$
can be arbitrarily transformed to a state with finite $\Delta$.

Finite $\mu$ would act on the free energy as an external field
 tending to make the diquark condensation favored. Consequently
even an infinitesimal introduction of $\mu$ leads to the diquark
condensation phase with zero chiral condensate in the chiral limit.
This is a peculiar feature of 2-color QCD and is in a sharp contrast
to the 3-color QCD. In fact, we can prove that  the minimizing
condition for our $F_{\mathrm{eff}}[\sigma,\Delta]$ does not allow
non-zero $\sigma$ at any $T$ and $\mu$ in the chiral limit. See
Appendix~\ref{app:proof} for the proof.

Taking the fact that $\sigma$ vanishes in the chiral limit for
granted, the free energy is simply written as
\begin{align}
 F_{\mathrm{eff}}[\Delta] =\frac{d}{2}|\Delta|^2-T\log\left\{
  1+4\cosh^2\left(E_0/T\right)\right\}\,,\qquad
 E_0=\mathrm{arccosh}\left(\sqrt{(d/2)^2|\Delta|^2+\cosh^2\mu}
  \right)\,.
\label{eq:free_energy_zero}
\end{align}
Assuming that the finite $T$ phase transition with fixed $\mu$ is of
second order, which will be confirmed numerically later, the critical
temperature $T_{\mathrm{c}}$ may be determined by expanding
$F_{\mathrm{eff}}[\Delta]$ in terms of $|\Delta|^2$ and extracting the
point where the coefficient of $|\Delta|^2$ changes its sign. Since
the expansion reads
\begin{align}
  F_{\mathrm{eff}}[\Delta]
   =-T\log\left\{3+2\cosh\left(2\mu/T\right)\right\}
   +\left\{\frac{d}{2}-\frac{d^2}{3+2\cosh\left(2\mu/T\right)}
   \frac{\sinh\left(2\mu/T\right)}{\sinh2\mu}\right\}|\Delta|^2
  +\mathrm{O}\left(|\Delta|^4\right)\,,
\end{align}
one finds  
\begin{align}
 T_{\mathrm{c}}(\mu)=2\mu\left\{\mathrm{arccosh}\left(
  \frac{3\sinh^2 2\mu+d\sqrt{4d^2+5\sinh^2 2\mu}}{2d^2-2\sinh^2 2\mu}
  \right)\right\}^{-1}\,.
\label{eq:critical_temp}
\end{align}
The values of $T_{\mathrm{c}}(\mu)$ for two typical cases  are
\begin{align}
 T_{\mathrm{c}}(\mu=0)=\frac{6}{5}\,, \qquad
 T_{\mathrm{c}}(\mu=\mu_{\mathrm{c}}=\frac{1}{2}\ \mathrm{arcsinh}\,d
  \simeq 0.91)=0\,
\label{eq:critical_temp_spec}
\end{align}
for $d=3$. $T_c$ is a monotonically decreasing function of $\mu$
connecting the above two points. This will be discussed later in
Sec.~\ref{sec:numerical} together with the case at finite $m$.

It is interesting to calculate the quark number density $\rho$
which is defined as
\begin{align}
 \rho(\mu,T)=-\frac{\partial F_{\mathrm{eff}}}{\partial\mu}
  =\frac{8\cosh\left(E_0/T\right)\sinh\left(E_0/T\right)}
   {1+4\cosh^2\left(E_0/T\right)}\cdot
   \frac{\partial E_0}{\partial\mu}\,,
\label{eq:density}
\end{align}
where $E_0$ is the quasi-quark mass in the chiral limit defined in 
Eq.~(\ref{eq:free_energy_zero}). Although the above expression seems
to be a little complicated, it can be simplified  by using the gap
equation;
\begin{align}
 0=\frac{\partial F_{\mathrm{eff}}}{\partial\Delta^\ast}
  =\frac{d}{2}\Delta-\frac{8\cosh\left(E_0/T\right)
  \sinh\left(E_0/T\right)}{1+4\cosh^2\left(E_0/T\right)}
  \cdot\frac{(d/2)^2\Delta}{2\cosh\mu\sinh\mu}\cdot
  \frac{\partial E_0}{\partial\mu}\,.
\label{eq:gap_eq}
\end{align}
In general the gap equation may have two solutions, $\Delta=0$ and
$\Delta\neq0$. The former (the latter) is the solution to minimize the
free energy for $T\ge T_{\mathrm{c}}(\mu)$ ($T<T_{\mathrm{c}}(\mu)$).
It can be alternatively said that the former (the latter) is the
solution for $\mu\ge \mu_{\mathrm{c}}(T)$
($\mu\le\mu_{\mathrm{c}}(T)$) with the critical chemical potential
$\mu_{\mathrm{c}}$ as a solution of Eq.~(\ref{eq:critical_temp}) in
terms of $\mu$.

By eliminating most of the complicated part of Eq.~(\ref{eq:density})
by means of the gap equation (\ref{eq:gap_eq}), we have the following
expressions;
\begin{align}
 \rho(\mu,T)=
 \begin{cases}
  \quad\left(2\sinh2\mu\right)/d\frac{}{}
  &\text{for}\quad \mu<\mu_{\mathrm{c}}(T)\quad (\Delta\neq0)\,,\\
  \displaystyle
  \,\frac{4\sinh\left(2\mu/T\right)}{3+2\cosh\left(2\mu/T\right)}
  &\text{for}\quad \mu\geq \mu_{\mathrm{c}}(T)\quad (\Delta=0)\,.
 \end{cases}
\label{eq:quark_density}
\end{align} 

At zero temperature, Eq.~(\ref{eq:quark_density}) is reduced to
\begin{align}
 \rho(\mu,T=0)=
 \begin{cases}
  \,\left(2\sinh2\mu\right)/d
  &\text{for}\quad \mu<\mu_{\mathrm{c}}(T=0)\,,\\
  \qquad 2\frac{}{}
  &\text{for}\quad \mu\geq\mu_{\mathrm{c}}(T=0)\,
 \end{cases}
 \label{eq:quark_density_T0}
\end{align}
with
\begin{align}
\mu_{\mathrm{c}}(T=0)=\frac{1}{2}\mathrm{arcsinh}\,d ,
\label{eq:muc0}
\end{align}
at which the quark number density $\rho$ gets to be saturated to
two. Note that quarks on each lattice site have only two degrees of
freedom in the color space because the spin and flavor degrees of
freedom are dispersed on the staggered lattice sites. Therefore ``two''
is the maximum number of quarks placed on each lattice site due to the
Pauli exclusion principle.


\subsection{Case at  Zero Temperature with Finite $m$}
\label{T0mn0}

Even a small quark mass could modify the phase structure substantially
from that given in Sec.~\ref{sec:analytic-A}. This is because the
pion, that is the Nambu-Goldstone mode associated with the chiral
$\mathrm{U_A(1)}$ symmetry breaking, comes to have a finite mass,
$m_\pi \propto m^{1/2}$. As a result, as far as the quark chemical
potential is smaller than a threshold value, the vacuum is empty and
the diquark condensate vanishes; $\rho(\mu,T=0)=0$ and
$\Delta(\mu, T=0)=0$.
 
To clarify such threshold effects in more detail, let us now focus our
attention on the $T=0$ system with finite $m$. We derive analytic
formulae for lower-critical chemical potential
$\mu_{\mathrm{c}}^{\mathrm{low}}$ at which $\Delta$ starts to be
non-vanishing and the upper-critical chemical potential
$\mu_{\mathrm{c}}^{\mathrm{up}}$ at which $\Delta$ ceases to be
non-vanishing.

The free energy at $T=0$ with finite $m$ takes a simple form;
\begin{align}
 F_{\mathrm{eff}}[\sigma,\Delta]
  =\frac{d}{2}\sigma^2+\frac{d}{2}|\Delta|^2-\left(E_++E_-\right)\,.
\end{align}
The expansion of $F_{\mathrm{eff}}[\sigma,\Delta]$ in terms of
$|\Delta|^2$ near the threshold gives a condition to determine the
critical chemical potential $\mu_{\mathrm{c}}$,
\begin{align}
 \frac{d}{2}-\frac{d^2}{8\cosh E\cosh\mu_{\mathrm{c}}}\left(
  \frac{1}{\sinh(E+\mu_{\mathrm{c}})}+\frac{1}
  {|\sinh(E-\mu_{\mathrm{c}})|}\right)=0\,,
\label{eq:sta_chem}
\end{align}
where $\textstyle E=\mathrm{arccosh}\left(\sqrt{1+M^2}\right)$ and 
$M=m+d\sigma/2$ as defined in Eq.~(\ref{eq:d_mass}). In
Eq.~(\ref{eq:sta_chem}) the dynamical mass $M$, or $\sigma$, is
determined by the condition to minimize the free energy at the
threshold. We can reduce that free energy much more by putting
$\Delta=0$,
\begin{align}
 F_{\mathrm{eff}}[\sigma] = \frac{d}{2}\sigma^2
  -\left(E+\mu+|E-\mu|\right)\,.
\label{eq:free_energy_muc}
 \end{align}
The stationary condition of the free energy $F_{\mathrm{eff}}[\sigma]$
with respect to $\sigma$ yields the following chiral gap equations;
\begin{align}
 \frac{2}{d}(M-m)=
  \begin{cases}
   \, (1+M^2)^{-1/2}  &\text{for}\quad \mu<E\,,\\
   \qquad0\frac{}{}&\text{for}\quad \mu\ge E\,.
  \end{cases}
\label{eq:gap_eq_chiral}
\end{align}
Combining Eq.~(\ref{eq:sta_chem}) and Eq.~(\ref{eq:gap_eq_chiral}), we
find
\begin{align}
 \mu_{\mathrm{c}}=
  \begin{cases}
   \,\mu_{\mathrm{c}}^{\mathrm{low}}
   = \mathrm{arccosh}\sqrt{1+mM}
   &\text{for}\quad \mu<E\,,\\
   \,\mu_{\mathrm{c}}^{\mathrm{up}}
   = \mathrm{arccosh}\sqrt{1+K^2}
   &\text{for}\quad \mu\ge E\,,\frac{}{}
  \end{cases}
\label{eq:muc-threshold}
\end{align}
where we define $K$ as the solution of the equation,
\begin{align}
 \frac{2}{d}\left(K-\frac{m^2}{K}\right)=(1+K^2)^{-1/2}\,.
\label{eq:gap_eq_k}
\end{align}

With finite $m$ at $T=0$, the empty vacuum gives $\rho=0$ and $\Delta=0$ 
as long as $\mu<\mu_{\mathrm{c}}^{\mathrm{low}}$. On the other hand, 
the non-vanishing value of $\Delta$ is possible for
$\mu_{\mathrm{c}}^{\mathrm{low}}\le\mu\le\mu_{\mathrm{c}}^{\mathrm{up}}$. 
For $\mu>\mu_{\mathrm{c}}^{\mathrm{up}}$, the saturation of the quark
number density  occurs leading to  $\rho=2$ and $\Delta=0$.
 These behaviors will be confirmed numerically in
Sec.~\ref{sec:numerical}.

Note that we have not assumed the quark mass $m$ to be small, which is
in contrast to the approaches based on chiral perturbation theory
\cite{ChPT0,ChPT}. Therefore, Eq.~(\ref{eq:muc-threshold}) can relate
the critical chemical potentials to arbitrary values of $m$. For
sufficiently small $m$, we have a relation;
\begin{align}
 \mu_{\mathrm{c}}^{\mathrm{low}} 
   = m^{1/2} \cdot \left\{\frac{(1+d^2)^{1/2}-1}{2}\right\}^{1/4}\,,
 \qquad
 \mu_{\mathrm{c}}^{\mathrm{up}}   = \frac{1}{2}\mathrm{arcsinh}\,d\,.
\end{align}
The former may be  rewritten as
$\mu_{\mathrm{c}}^{\mathrm{low}}=m_{\pi}/2$ with $m_{\pi}$
obtained from the excitation spectrum in the vacuum \cite{klu83} up to
the leading order of the $1/d$ expansion. This observation is
consistent with the discussion given in \cite{ChPT0}.
The latter relation is simply rewritten as
$\mu_{\mathrm{c}}^{\mathrm{up}}=\mu_{\mathrm{c}}(T=0)$ with
$\mu_{\mathrm{c}}(T=0)$ defined in Eq.(\ref{eq:muc0}).

For sufficiently large $m$, we find asymptotically,
\begin{align}
 \mu_{\mathrm{c}}^{\mathrm{low}}\simeq \mu_{\mathrm{c}}^{\mathrm{up}}
  \simeq \mathrm{arccosh}\,m\, .
\end{align}
This is because $M\simeq K\simeq m$ for large $m$ as is evident from
Eqs.~(\ref{eq:gap_eq_chiral}) and (\ref{eq:gap_eq_k}).
  

\section{NUMERICAL RESULTS ON THE PHASE STRUCTURE}
\label{sec:numerical}

In this section, we determine the chiral condensate $\sigma$ and the
diquark condensate $\Delta$ numerically by minimizing the effective
free energy in Eq.~(\ref{eq:free_energy}). The quark number density
$\rho$ is also calculated numerically. The results are shown in
Figs.~\ref{fig:massless} and \ref{fig:massive} for $m=0$ and $m=0.02$,
respectively. The phase diagrams of the system in the $T$-$\mu$ plane,
in the $\mu$-$m$ plane, and in the three dimensional $T$-$\mu$-$m$
space are also shown in Figs.~\ref{fig:phase_diagram},
\ref{fig:phase_diagram-MuM}, and \ref{fig:phase_diagram-3D},
respectively.

\subsection{Case for $m=0$}

\begin{figure}[bhtp]
 \begin{center}
  \includegraphics[width=16cm,clip]{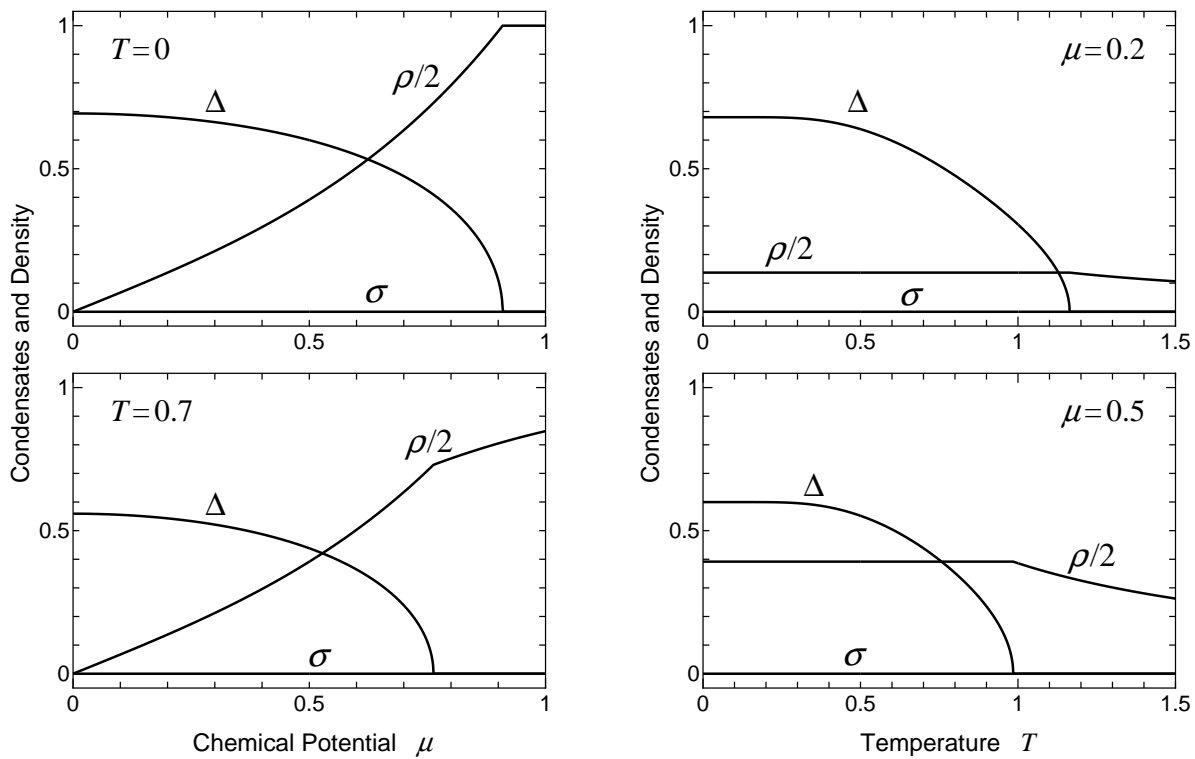}
  \caption{Chiral condensate $\sigma$, diquark condensate $\Delta$,
  and the quark number density $\rho$ for $m=0$ with $d=3$.
  In the left panels they are plotted as functions of
  the chemical potential $\mu$ for two typical values of temperature.
  In the right panels, they are plotted as functions of temperature $T$ 
  for two typical values of $\mu$.
  All  the dimensionful quantities are  in unit of the lattice
  spacing $a$,  which is implicitly understood in all the figures
  in this section.}
\label{fig:massless}
 \end{center}
\end{figure}

Let us first consider the diquark condensate and the quark number
density as functions of $\mu$ and $T$ shown in the four panels of
Fig.~\ref{fig:massless}. (Note that the chiral condensate is always
zero in the chiral limit as we have proved in
Appendix~\ref{app:proof}.)

At $T=0$ (the upper left panel), the diquark condensate $\Delta$ 
decreases monotonously as a function of $\mu$ and shows a second 
order transition when $\mu$ becomes of order unity. On the other hand,
the quark number density $\rho$ increases linearly for small $\mu$ and
grows more rapidly for large $\mu$ until the saturation point where
quarks occupy the maximally allowed configurations by the Fermi
statistics. (See the analytic formula given in
Eq.~(\ref{eq:quark_density_T0}).)  Those behaviors of $\Delta$ and
$\rho$ are also observed in the recent Monte-Carlo simulations of
2-color QCD \cite{kog02}. Note that the rapid increase of $\rho$ near
the upper-critical chemical potential $\mu_{\mathrm{c}}^{\mathrm{up}}$
takes place even in the strong coupling limit as shown here. Namely it
does not necessarily be an indication of the existence of free quarks
at high density unlike the suggestion given in the last reference in
\cite{kog02}.
     
As $T$ increases, the magnitude of the diquark condensate decreases
by the thermal excitations of quark and anti-quark pairs in the last
term of Eq.~(\ref{eq:free_energy}), which is shown in the lower left
panel of Fig.~\ref{fig:massless}. It is worth mentioning here that the
diquark condensate disappears even before the complete saturation
($\rho=2$) takes place.

Next we consider the diquark condensate as a functions of $T$ for two
typical values of the chemical potential in the right panels of
Fig.~\ref{fig:massless}. The diquark condensate shows a second order
transition at $T_{\mathrm{c}}$ given analytically by
Eq.~(\ref{eq:critical_temp}).

\subsection{Case for $m\neq 0$}

\begin{figure}[bhtp]
 \begin{center}
  \includegraphics[width=16cm,clip]{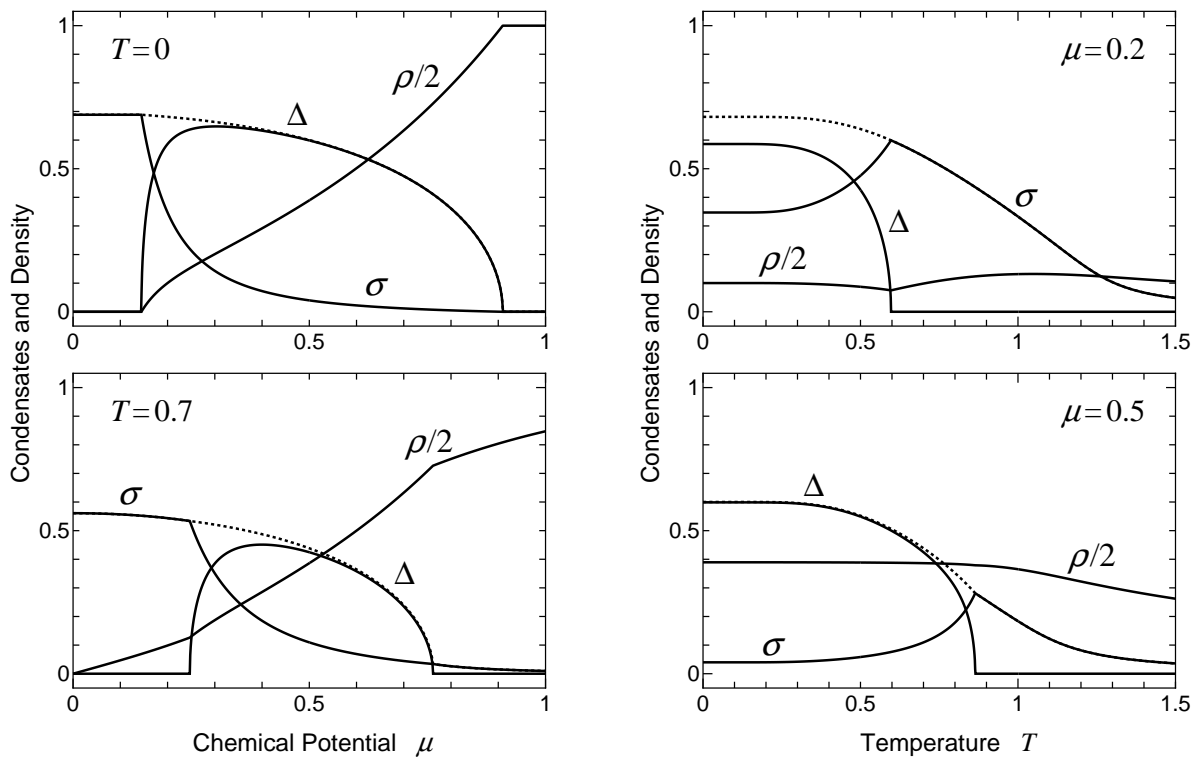}
  \caption{Chiral condensate $\sigma$, diquark condensate $\Delta$,
  and quark number density $\rho$ for $m=0.02$ with $d=3$.
  All the definitions are the same with those of Fig.~\ref{fig:massless}
  except that the dotted line indicates a total magnitude of
  the condensates $\sqrt{\sigma^2+\Delta^2}$.}
\label{fig:massive}
 \end{center}
\end{figure}

In the upper left panel of Fig.~\ref{fig:massive}, the chiral
condensate, the diquark condensate, and the quark number density are
shown as functions of $\mu$ for small quark mass $m=0.02$ at $T=0$. As
we have discussed in Sec.~\ref{T0mn0}, there exists a lower-critical
chemical potential $\mu_{\mathrm{c}}^{\mathrm{low}}$ given by
Eq. (\ref{eq:muc-threshold}). Both $\Delta$ and $\rho$ start to take
finite values only for $\mu > \mu_{\mathrm{c}}^{\mathrm{low}}$ at
$T=0$.

One can view the behavior of chiral and diquark condensates with
finite quark mass as the manifestation of two different mechanisms: One
is a continuous ``rotation'' from the chiral direction to the diquark
direction near $\mu=\mu_{\mathrm{c}}^{\mathrm{low}}$ or
$T=T_{\mathrm{c}}(\mu)$ with $\sqrt{\sigma^2+\Delta^2}$ (shown by the
dotted line) varying smoothly. The other is the saturation effect
which forces the diquark condensate to decrease and disappear
for large $\mu$ as seen in the previous case of $m=0$.

The ``rotation'' can be understood as follows: The free energy in the
mean field approximation at small $m$ and $\mu$ has an approximate
  symmetry which mixes the chiral condensate with the diquark one 
as we have discussed in Sec.~\ref{sec:analytic-A}. The effect of $m$
($\mu$) is to break this symmetry in the direction of the chiral
(diquark) condensation favored. Therefore, a relatively large chiral
condensate predominantly appears for small $\mu$ region.
\footnote{Note that the chiral symmetry is explicitly broken by $m$,
 thus the chiral condensate is always non-vanishing although it is
 suppressed in magnitude at high $T$ and $\mu$.}
Just above the lower-critical  chemical potential
$\mu_{\mathrm{c}}^{\mathrm{low}}$, the chiral condensate decreases,
while the diquark condensate increases because the effect of $\mu$
surpasses that of $m$. As $\mu$ becomes large, the diquark condensate
begins to decrease in turn by the effect of the saturation
and eventually disappears when $\mu$ exceeds the upper-critical value
$\mu_{\mathrm{c}}^{\mathrm{up}}$ (order of unity for $T=0$).
 
Similar ``rotation'' and the saturation effect are also seen at finite
$T$ as shown in the lower left panel of Fig.~\ref{fig:massive}. At
finite $T$, both the chiral and diquark condensates are suppressed due
to the effect of temperature and the diquark condensate disappears
\textit{before} the complete saturation occurs.

Next we consider the chiral and diquark condensates as functions of
$T$ (the right panels of Fig.~\ref{fig:massive}). At low $T$, both
the chiral and diquark condensates have finite values for $\mu =0.2$
and $0.5$. The diquark condensate decreases monotonously as $T$
increases and shows a second order transition. On the other hand, the
chiral condensate increases as the diquark condensate decreases so
that the total condensate $\sqrt{\sigma^2+\Delta^2}$ is a smoothly
varying function of $T$. The  understanding based on the
chiral-diquark mixing symmetry is thus valid. An interesting
observation is that the chiral condensate, although it is a continuous
function of $T$, has a cusp  shape associated with the phase
transition of the diquark condensate.

Finally let us compare the $m=0$ case in Fig.~\ref{fig:massless} and
the $m=0.02$ case in Fig.~\ref{fig:massive}. Looking into two figures
at the same temperature or chemical potential, we find that the
diquark condensate $\Delta$ for $m=0$ and the total condensate
$\sqrt{\sigma^2+\Delta^2}$ for $m=0.02$ have almost the same behavior.
This indicates that although the current quark mass suppresses the
diquark condensate, the price to pay is to increase the chiral
condensate so as to  make the total condensate insensitive to the
presence of small quark mass. Restating this by use of the radial and
the angle variables defined by $\lambda\sin\phi=\sigma$ and
$\lambda\cos\phi=|\Delta|$ as in Eq.~(\ref{eq:radial-angle}), a small
$m$ hardly changes the behavior of $\lambda$ but shifts $\phi$ from
zero.

\subsection{Phase Diagrams}

\begin{center} 
 \begin{figure}[bthp]
 \begin{minipage}[ct]{8.5cm}
 \begin{center}
  \includegraphics[width=8cm,clip]{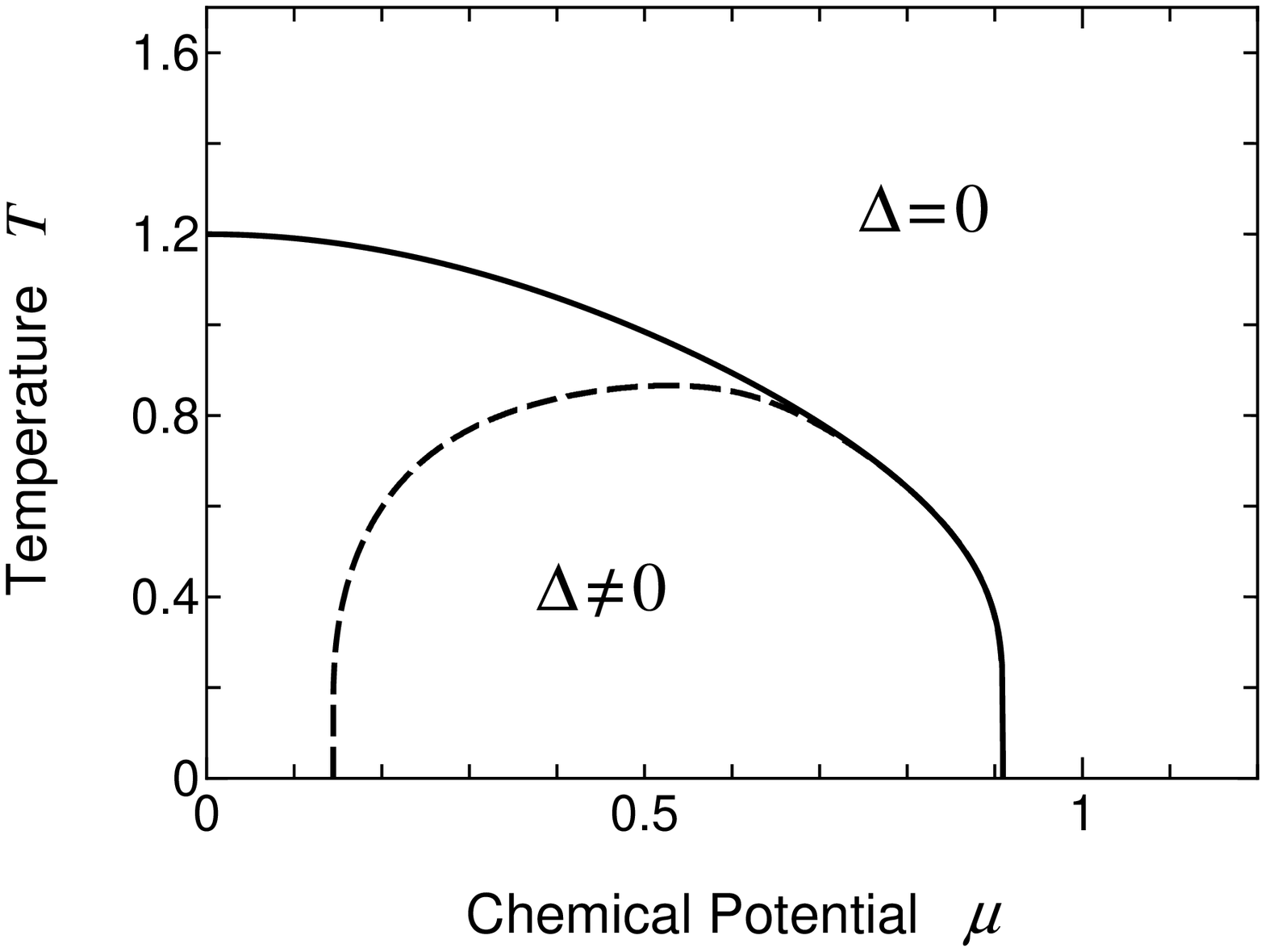}
  \caption{Phase diagram of the strong coupling 2-color QCD in the
  $T$-$\mu$ plane. Solid (dashed) line denotes the critical line for
  diquark condensation for $m=0$ ($m=0.02$).
 \label{fig:phase_diagram}}
 \end{center}
 \end{minipage}\hfill
 \begin{minipage}{8.5cm}
 \begin{center}
  \includegraphics[width=8cm,clip]{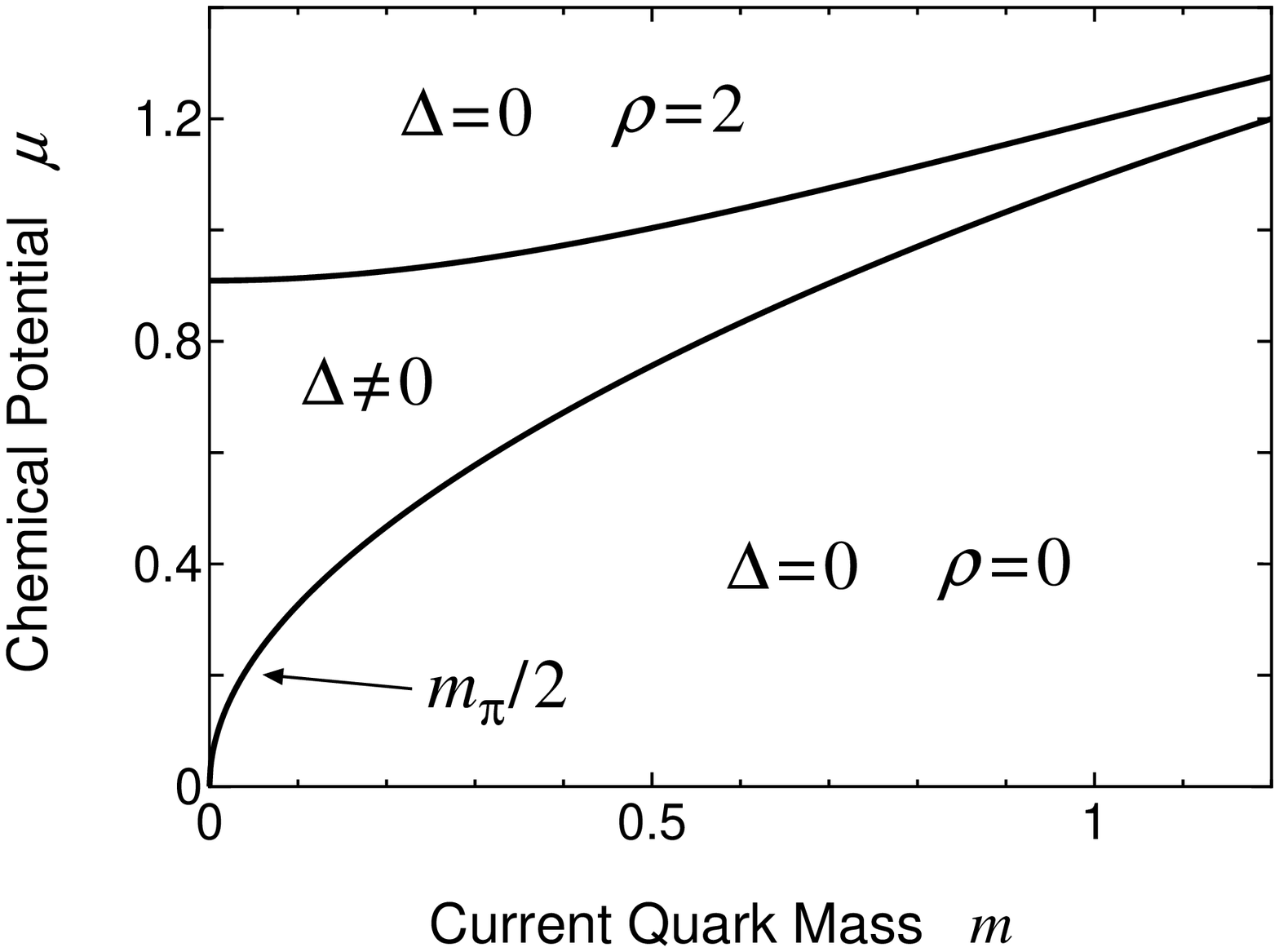}
  \caption{Phase diagram of strong coupling 2-color QCD in the
  $\mu$-$m$ plane at $T=0$. Two solid lines separate the region where
  $\Delta=0$ from the region where $\Delta\not=0$.
 \label{fig:phase_diagram-MuM}}
 \end{center}
 \end{minipage}
 \end{figure}
\end{center}
 
Now we show the phase diagram of the strong coupling 2-color QCD in
the $T$-$\mu$ plane in Fig.~\ref{fig:phase_diagram}. The solid line
denotes a critical line separating diquark superfluid phase
$\Delta\not=0$ and the normal phase $\Delta=0$ in the chiral limit
$m=0$, which is determined analytically by
Eq.~(\ref{eq:critical_temp}). The dashed line denotes the critical
line determined numerically for $m=0.02$. The phase transition is of
second order on these critical lines. The chiral condensate $\sigma$
is everywhere zero for $m=0$, while it is everywhere finite for
$m\neq 0$. In the latter case, however, $\sigma$ is particularly large
in the region between the solid line and the dashed line. Shown in
Fig.~\ref{fig:phase_diagram-MuM} is a phase diagram in $\mu$-$m$ plane
at $T=0$. The lower right of the figure corresponds to the vacuum with
no baryon number present, $\rho=0$. On the other hand, the upper left
of the figure corresponds to the saturated system, $\rho=2$, in which
every lattice site is occupied by two quarks. There is a region
with $0<\rho<2$ and $\Delta\not=0$ bounded by the above two limiting
cases, which is separated by two critical lines given in
Eq.(\ref{eq:muc-threshold}).

\begin{figure}[tbph]
 \begin{center}
  \includegraphics[width=10cm,clip]{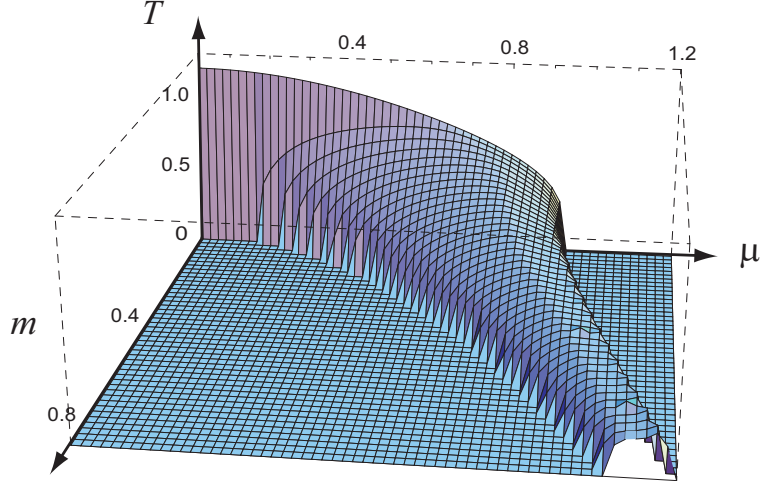}
  \caption{Phase structure of strong coupling 2-color QCD in the
  $T$-$\mu$-$m$ space. The surface  represents the critical surface
  for the diquark condensation, which separates the region where
  $\Delta=0$ from the region where $\Delta\not=0$.
  \label{fig:phase_diagram-3D}}
 \end{center}
\end{figure}

Finally,  bringing all the discussions together, the phase structure
in the three dimensional $T$-$\mu$-$m$ space is shown in
Fig.~\ref{fig:phase_diagram-3D}. The diquark condensate has a
none-vanishing value inside the critical surface and the phase
transition is of second order everywhere on this critical surface.
 The second order phase transition is consistent with other analyses
employing the mean field approximation; the random matrix model and  
the Nambu$-$Jona-Lasinio model \cite{RMT}, and also the chiral
perturbation theory at $T=0$ \cite{ChPT0}. On the other hand,
Monte-Carlo simulations of 2-color QCD \cite{kog02} show indication of
a tricritical point in the $T$-$\mu$ plane at which the property of
the critical line changes from the second order to the first order as
$\mu$ increases. This is also supported by the chiral perturbation
theory beyond the mean field approximation at finite $T$ \cite{ChPT}.

Aside from the fact that we are working in the strong coupling limit
($g\rightarrow \infty$), though the lattice simulations are aiming at
the weak coupling limit ($g\rightarrow 0$), it would be of great
interests to go beyond the mean field approximation in our analysis
and study corrections to the phase structure given in
Fig.~\ref{fig:phase_diagram-3D}.

\section{SUMMARY AND CONCLUDING REMARKS}
\label{sec:summary}

In this paper, we have studied the phase structure of 2-color QCD in
the strong coupling limit formulated on the lattice. We have employed
the $1/d$ expansion (only in the spatial direction) and the mean field
approximation to derive the free energy written in terms of the chiral
condensate $\sigma$  and the diquark condensate $\Delta$ at finite
$T$,  $\mu$ and $m$. A major advantage of our approach is that we can
derive rather simple formulae in an analytic way for the critical
temperature ($T_{\mathrm{c}}$), the critical chemical potentials
($\mu_{\mathrm{c}}^{\mathrm{low}}$ and
$\mu_{\mathrm{c}}^{\mathrm{up}}$), and the quark number density
($\rho$), which are useful to have physical insight into the problem.
Although our results are in principle limited in the strong coupling,
the behavior of $\sigma$, $\Delta$, and $\rho$ in the $T$-$\mu$-$m$
space has remarkable qualitative agreement with the recent lattice
data. Since we do not have to rely on any assumption of small $\mu$
nor small $m$ in our approach, the strong coupling analysis presented
in this paper is complementary to the results of the chiral
perturbation theory and directly provides a useful guide to the lattice
2color-QCD simulations.
 
There are several directions worth to be explored starting from the
present work. Among others, the derivation  of an effective
action written not only with the chiral and diquark fields but also
with the Polyakov loop, $L(x)$, is interesting  because this will 
allow us a full comparison for all the physical quantities calculated in
our formalism and lattice simulations \cite{kog02}. This is a
straightforward generalization of the works in \cite{fuku03}, which
allows us to analyze $\langle L(x)\rangle$ under the influence of
chiral and diquark condensates. 
 
Also calculation of the meson and diquark spectra including
pseudo-scalar and pseudo-diquark channels will give us deeper
understanding about the non-static feature of the present system. 
Another direction is to study the response of the Bose liquid
discussed in this paper under external fields. Although mesons and
diquarks are color neutral in 2-color QCD, they can have electric
charge so that the external electromagnetic field leads to a
non-trivial change of the phase structure if the field intensity is
enough strong. Generalization of the whole machinery in this paper to
strong coupling 3-color QCD is also one of the challenging problems to
be studied in the future.

\begin{acknowledgments}
Authors are grateful to K.~Iida, M.~Tachibana, S.~Sasaki, H.~Abuki,
and K.~Itakura for continuous and stimulating discussions on 2-color
QCD. T.H.\ thanks G.~Baym for discussions on the Bose-Einstein
condensation of tightly bound diquarks. This work is partially
supported by the Grants-in-Aid of the Japanese Ministry of Education,
Science and Culture (No.~15540254). K.~F.\ is supported by the Japan
Society for the Promotion of Science for Young Scientists.
\end{acknowledgments}
 
\newpage


\appendix


\section{SUMMATION OVER THE MATSUBARA FREQUENCY}
\label{app:summation}

Let us first take the logarithm of Eq.(\ref{eq:pre-det});
\begin{align}
 \log\sqrt{\mathrm{Det}[G^{-1}]}
  &=\sum_{m=1}^{N_\tau}\log\left[-\cos^2\left(k_m+\theta/N_\tau\right)
  +2Y\cos\left(k_m+\theta/N_\tau\right)+X^2\right]\,.
\end{align}
Here $X$ and $Y$ are defined as
\begin{align}
 X^2=\cosh^2\mu+M^2+\left(\frac{d}{2}\right)^2|\Delta|^2\,,\qquad
 Y=M\sinh\mu\,.
\end{align}
By differentiating $\log\sqrt{\mathrm{Det}[G^{-1}]}$ with respect to
$X$, we obtain
\begin{align}
 \frac{\partial}{\partial X}\log\sqrt{\mathrm{Det}[G^{-1}]}
  =\frac X{\sqrt{X^2+Y^2}}\sum_{m=1}^{N_\tau}
 \left[\frac1{\cos\left(k_m+\theta/N_\tau\right)-Y+\sqrt{X^2+Y^2}}
 -\frac1{\cos\left(k_m+\theta/N_\tau\right)-Y-\sqrt{X^2+Y^2}}\right]\,.
\end{align}
Because $\cos(k_m+\theta/N_\tau)$ is invariant under the shift
$m\to m+N_\tau$, we can make the summation  of $m$ over the range
$m=-\infty$ to $m=+\infty$ with an appropriate degeneracy factor
$\Omega$. Then the  residue theorem enables us to replace the
summation by a complex integral as
\begin{align}
 \begin{split}
  \frac\partial{\partial{X}}\log\sqrt{\mathrm{Det}[G^{-1}]}
  =\frac{X}{\sqrt{X^2+Y^2}}\frac{1}{\Omega}
 &\left[\oint\frac{\mathrm{d}z}{2\pi\mathrm{i}}
  \frac{1}{\cos\left(z+\theta/N_\tau\right)+k_-}
  \frac{-\mathrm{i}N_\tau}{1+\mathrm{e}^{\mathrm{i}N_\tau z}}
  -\sum_{\bar{z}}\frac{1}{-\sin\bar{z}}
  \frac{-\mathrm{i}N_\tau}{1+\mathrm{e}^{\mathrm{i}N_\tau\bar{z}}}
  \right.\\
 &\left.-\oint\frac{\mathrm{d}w}{2\pi\mathrm{i}}
  \frac{1}{\cos\left(w+\theta/N_\tau\right)-k_+}
  \frac{-\mathrm{i}N_\tau}{1+\mathrm{e}^{\mathrm{i}N_\tau w}}
  +\sum_{\bar{w}}\frac{1}{-\sin\bar{w}}
  \frac{-\mathrm{i}N_\tau}{1+\mathrm{e}^{\mathrm{i}N_\tau\bar{w}}}
  \right]\,,
 \end{split}
\label{eq:problem}
\end{align}
where
\begin{align}
 k_\pm=\sqrt{X^2+Y^2}\pm Y\,.
\end{align}
Owing to the infinite range of the summation over $m$, we can choose
the closed contour at infinity for the complex integrals with respect
to $z$ and $w$, and thus such complex integrals go to zero. $\bar{z}$
and $\bar{w}$ are the residues satisfying
\begin{align}
 \cos\left(\bar{z}+\theta/N_\tau\right)+k_-=0\,,
 \qquad\cos\left(\bar{w}+\theta/N_\tau\right)-k_+=0\,.
\end{align}
Solving these equations, we obtain
\begin{align}
 \bar z+\theta/N_\tau=\pm\mathrm iE_-+\pi+2\pi n\,,\qquad
 \bar w+\theta/N_\tau=\pm\mathrm iE_++2\pi n\,,\qquad\quad
 (n=-\infty,\dots,\infty)\,,
\label{eq:solution}
\end{align}
where
\begin{align}
 E_\pm=\mathrm{arccosh}k_\pm\,.
\label{eq:e_pm}
\end{align}
Substituting Eq.~(\ref{eq:solution}) into Eq.~(\ref{eq:problem}), we
obtain
\begin{align}
 &\frac\partial{\partial{X}}\log\sqrt{\mathrm{Det}[G^{-1}]} \notag\\
 &=\frac{X}{\sqrt{X^2+Y^2}}\frac{1}{\Omega}\sum_{n=-\infty}^\infty
  \left[\frac{1}{-\mathrm{i}\sqrt{k_-^2-1}}\left(
  \frac{-\mathrm{i}N_\tau}{1+\mathrm{e}^{-N_\tau E_--\mathrm{i}\theta
  +\mathrm{i}N_\tau\pi}}-
  \frac{-\mathrm{i}N_\tau}{1+\mathrm{e}^{N_\tau E_--\mathrm{i}\theta
  +\mathrm{i}N_\tau\pi}}\right)\right. \notag\\
 &\qquad\qquad\qquad\qquad\qquad\qquad\qquad\left.
  -\frac{1}{\mathrm{i}\sqrt{k_+^2-1}}\left(
  \frac{-\mathrm{i}N_\tau}{1+\mathrm{e}^{-N_\tau E_+-\mathrm{i}\theta}}
  -\frac{-\mathrm{i}N_\tau}{1+\mathrm{e}^{N_\tau E_+-\mathrm{i}\theta}}
  \right)\right]\notag\\
 &=\frac{\partial E_-}{\partial X}
  \left\{\frac{N_\tau}{1+\mathrm{e}^{-N_\tau E_--\mathrm{i}\theta}}
  -\frac{N_\tau}{1+\mathrm{e}^{N_\tau E_--\mathrm{i}\theta}}\right\}
  +\frac{\partial E_+}{\partial X}
  \left\{\frac{N_\tau}{1+\mathrm{e}^{-N_\tau E_+-\mathrm{i}\theta}}
  -\frac{N_\tau}{1+\mathrm{e}^{N_\tau E_+-\mathrm{i}\theta}}\right\}
  \notag\\
 &=\frac{\partial}{\partial X}
  \left[\log\left\{\mathrm{e}^{N_\tau E_-}
   +\mathrm{e}^{-\mathrm{i}\theta}\right\}
  +\log\left\{\mathrm{e}^{-N_\tau E_-}
   +\mathrm{e}^{-\mathrm{i}\theta}\right\}
  +\log\left\{\mathrm{e}^{N_\tau E_+}
   +\mathrm{e}^{-\mathrm{i}\theta}\right\}
  +\log\left\{\mathrm{e}^{-N_\tau E_+}
   +\mathrm{e}^{-\mathrm{i}\theta}\right\}\right]\notag\\
 &=\frac{\partial}{\partial X}
  \left[-2\mathrm i\theta+2\log2
   +\log\left\{\cos\theta+\cosh N_\tau E_-\right\}
   +\log\left\{\cos\theta+\cosh N_\tau E_+\right\}\right]\,.
\end{align}
We note that the degeneracy factor $\Omega$ is just canceled by the
infinite degeneracy of the summation on $n$. Also we have used the
fact that $N_\tau$ must be an even integer for the staggered fermion.
After the integration with respect to $X$, we find that
$\sqrt{\mathrm{Det}[G^{-1}]}$ is expressed in a rather simple form 
up to irrelevant factors,
\begin{align}
 \sqrt{\mathrm{Det}\left[G^{-1}_{ab}(m,n;\theta)\right]}
  =\left(\cos\theta+\cosh N_\tau E_-\right)\cdot
  \left(\cos\theta+\cosh N_\tau E_+\right)\,.
\label{eq:sqrtdet2}
\end{align}
This is the final form shown in Eq.~(\ref{eq:sqrtdet}) in the text.


\section{PROOF OF $\sigma=0$ AT THE GLOBAL MINIMUM OF
 $F_{\mathrm{eff}}[\sigma,\Delta]|_{m=0}$}
\label{app:proof}

For $m=0$, it is useful to rewrite the free energy
Eq.~(\ref{eq:free_energy}) in terms of the radial and angle variables,
\begin{align}
 \sigma=\lambda\sin\phi\,,\qquad
 |\Delta|=\lambda\cos\phi\,.
\label{eq:radial-angle}
\end{align}
Then the free energy is
\begin{align}
 F_{\mathrm{eff}}[\lambda,\phi]=\frac{d}{2}\lambda^2-T\log\left\{
  1+4\cosh\left(E_+/T\right)\cdot\cosh\left(E_-/T\right)\right\}\,,
\end{align}
with the quasi-quark energy given by
\begin{align}
 E_\pm=\mathrm{arccosh}\left(\sqrt{(d/2)^2\lambda^2+\cosh^2\mu
  +(d/2)^2\lambda^2\sinh^2\mu\sin^2\phi}\pm\frac{d}{2}\lambda
  \sinh\mu\sin\phi\right)\,.
\end{align}
What we are going to prove here is that
$F_{\mathrm{eff}}[\lambda,\phi]$ has the global minimum at $\phi=0$
for arbitrary $T$, $\mu$, and $\lambda$. Actually we can prove the
following statement in a more abstract expression;
\begin{align*}
 &\cosh\{a\,\mathrm{arccosh}(b\,\mathrm{e}^c)\}\cdot
  \cosh\{a\,\mathrm{arccosh}(b\,\mathrm{e}^{-c})\}
  \text{\, has the global maximum at $c=0$}\,,\\
 &\text{where } a>1,\text{ } b>1, \text{ and } -\log b<c<\log b\,.
\end{align*}
We can apply this corollary to our problem to prove that $\phi=0$ is
the global minimum of the free energy. Obviously $\phi=0$ turns out to
be the global maximum of $\cosh(E_+/T)\cdot\cosh(E_-/T)$ once we
substitute,
\begin{align}
 a=1/T,\quad b=\sqrt{(d/2)^2\lambda^2+\cosh^2\mu},\quad
 c=\log\left[\sqrt{1+(d\lambda/2b)^2\sinh^2\mu\sin^2\phi}
  +(d\lambda/2b)\sinh\mu\sin\phi\right]\,.
\end{align}
Thus we have proven that $F_{\mathrm{eff}}[\lambda,\phi]$ is globally
minimized at $\phi=0$, in other words, the chiral condensate vanishes
in the chiral limit.

\newpage


\end{document}